\documentclass[jpb,preprint]{revtex4}%
\usepackage{amsfonts}
\usepackage{amsmath}
\usepackage{amssymb}
\usepackage{graphicx}%
\begin{document}
\title{Generation of frequency entanglement by rotating Doppler effect}
\author{Bolong Yi}
\author{Ling Chen}
\thanks{Corresponding author: lingchen@cug.edu.cn}
\author{Baocheng Zhang}
\thanks{Corresponding author: zhangbaocheng@cug.edu.cn}
\affiliation{School of Mathematics and Physics, China University of Geosciences, Wuhan
430074, China}
\keywords{frequency, entanglement, rotating Doppler effect}
\begin{abstract}
We propose a method to generate the frequency entanglement, allowing a
continuous generation of entangled two-photon states in a hybrid degree of
freedom by post-manipulation. Our method is based on type-II spontaneous
parametric down-conversion in a nonlinear crystal and the rotation Doppler
effect by rotating the q-plates, without preset discrete frequency
entanglement. This allows the arbitrary modification of frequency entangled
photons in a wide frequency range at room temperature, offering enhanced
flexibility for quantum information tasks and quantum metrology. We also
analyze the entanglement state by a combined calculation for the joint
spectrum and Hong-Ou-Mandel interference of the two photons, which can be used
to reconstruct a restricted density matrix in the frequency space.

\end{abstract}
\maketitle



\section{Introduction}

Entanglement \cite{hhh2009} exhibits phenomena that are fundamentally distinct
from classical physics. The states of entangled particles are intricately
correlated, such that even when separated by vast distances, measuring the
state of one particle instantaneously affects the state of the other,
demonstrating a non-local correlation that transcends spatial separation. This
phenomenon challenges traditional notions of causality and the transmission of
matter, revealing the counterintuitive nature of the quantum world.
Furthermore, entanglement enables the \textquotedblleft
teleportation\textquotedblright\ \cite{bbcw1993} of quantum information
without physical transmission, providing a critical foundation for
cutting-edge technologies such as quantum computation and quantum communication.

Entanglement can utilize a wide range of degrees of freedom, including
polarization, spin angular momentum, orbital angular momentum, and more
\cite{wia2015,fss2018}. Frequency is one of the most common degrees of freedom
for entanglement. While frequency is inherently a continuous degree of freedom
and can be used to encode information \cite{mrd2016,drm2018,adb2018}, it can
also be treated as a discrete degree of freedom when divided into specific
frequency bins \cite{cne2010,krl2019}. There are many methods to generate
frequency entanglement, such as using nonlinear crystals to produce photons
with different frequencies \cite{kmw1995}, electronic and atomic energy level
transitions to create entangled sources with distinct frequencies
\cite{hbd2015}, and phonon vibrational modes in solid-state systems to
generate frequency entanglement \cite{bsz2019}. Photons, as a source of
frequency entanglement, offer several advantages, including low propagation
loss over long distances in optical fibers, ease of manipulation using
inexpensive optical components, long coherence times with minimal
environmental interference, and mature detection technologies. Such entangled
states have significant applications in quantum communication, quantum key
distribution (QKD), and quantum sensing. For instance, frequency entanglement
enables high-dimensional quantum teleportation and high-capacity quantum
communication protocols, significantly enhancing channel capacity and
information transmission efficiency \cite{brs2015}. Additionally,
frequency-entangled states exhibit strong robustness against channel noise and
interference, making them particularly suitable for long-distance quantum
communication \cite{llp2017}. In the field of quantum sensing, frequency
entanglement is utilized for ultra-high-precision spectral measurements and
frequency standard calibration, demonstrating its potential in precision
metrology \cite{adb2018}.

In this paper, we propose a method to generate the frequency entangled states
based on the spontaneous parametric down-conversion (SPDC) using a $\beta
$-barium borate (BBO) crystal \cite{phj1994,mw1995} and the rotating Doppler
effect (RDE) \cite{bag1981,bb1997,crp1998} using a rotating q-plate. It is
known that the photons generated by BBO are purely polarization-entangled and
lack a frequency difference between the two photon states. To construct a
frequency entangled state, we first convert the linearly polarized entangled
photons into circularly polarized entangled photons using a quarter-wave plate
(QWP). Then, a rotating q-plate is used to impart torque to the photons
\cite{pja1966}, inducing a frequency shift by the rotating Doppler effect of
light \cite{ga1979}. Subsequently, a polarizer is employed to eliminate the
polarization influence, thereby constructing a purely frequency-entangled
state. The entanglement can be verified using Hong-Ou-Mandel (HOM)
interference \cite{hom1987}, and the rotational speed of the wave plate can be
determined based on the interference pattern.

\section{Method of entanglement generation}

As a kind of the carriers of quantum information, photons provide a large
variety of degrees of freedom, such as frequency, polarization, spatial modes,
linear, spin or orbital angular momentum \cite{wia2015,fss2018}. As
well-known, frequency entanglement and hybrid polarization-frequency
entanglement \cite{frl2023} between photon pairs arises naturally in SPDC
process as a consequence of energy conservation under the phase-matching
condition. However, continuous manipulation of the frequency entanglement
associated with spatial mode (or orbital angular momentum, OAM) has not been
proposed so far. Here, we will propose a method to generate the frequency
entanglement using a rotating continuous manipulation, as presented in Fig. 1.
The first step is to generate a pair of polarization-entangled photons with
equal wavelength (frequency) directly from two intersections on two overlap
cones emitted by the type-II SPDC pumped in a nonlinear BBO crystal
\cite{kmw1995,clw2012}. This photon pair can be essentially described as an
entangled state expressed as%
\begin{equation}
\psi_{p}=\frac{1}{\sqrt{2}}\left(  \left\vert H\right\rangle \left\vert
V\right\rangle +\left\vert V\right\rangle \left\vert H\right\rangle \right)
\left\vert \omega\right\rangle , \label{ps}%
\end{equation}
where $\left\vert H\right\rangle $ ($\left\vert V\right\rangle $) represents
the horizontal (vertical) polarized state, and $\omega$ is the light
frequency. In particular, the phase difference from the crystal birefringence
and an overall phase shift are omitted in the expression (\ref{ps}). Such
states have been extensively applied in quantum communication and quantum
cryptography \cite{fap2010}.

\begin{figure}[ptb]
\centering
\includegraphics[width=0.9\columnwidth]{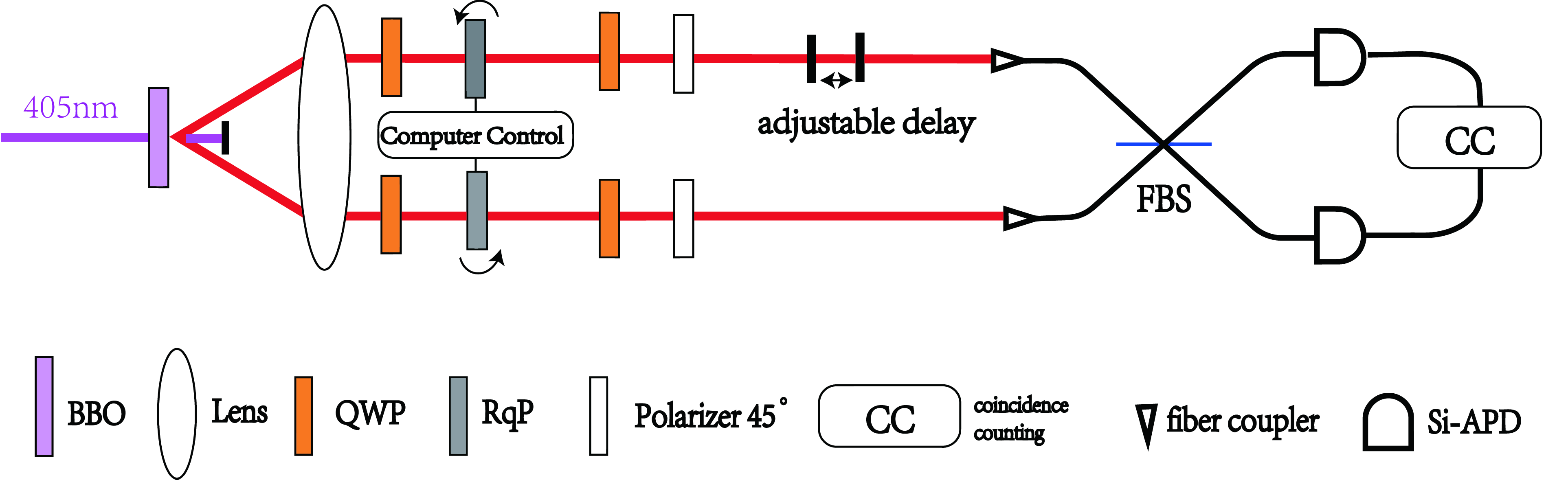}\caption{(Color online)
Schematic diagram of the experimental proposal. BBO: $\beta$-barium borate;
QWP: quarter-wave plate; RqP: rotating q-plate; APD: avalanche photodiode;
FBS: fiber beam-splitter.}%
\label{Fig1}%
\end{figure}

The polarized states can be easily transferred to spin angular momentum (SAM)
states, which is another degree of freedom of light with circular polarization
by passing through a QWP with a consistent wavelength. In Fig. 1, we employ
two identical QWPs to transfer this state (\ref{ps}) to the entangled state
between the SAM of photons which can be written as
\begin{equation}
\psi_{s}=\frac{1}{\sqrt{2}}\left(  \left\vert \sigma^{+}\right\rangle
\left\vert \sigma^{-}\right\rangle +\left\vert \sigma^{-}\right\rangle
\left\vert \sigma^{+}\right\rangle \right)  \left\vert \omega\right\rangle ,
\end{equation}
where $\sigma^{+}(\sigma^{-})=+1(-1)$ stands for left (right) circular polarizations.

More recently, it was achieved that the OAM is another degree of freedom of
the photons which can be efficiently produced by a meta-surface \cite{sfm2018}%
. An exquisitely designed meta-surface can change the photon's spin to hybrid
SAM-OAM entanglement. Ordinarily, the entanglement could be realized in two
SPDC photons with the same frequency $\omega$ by post-selection or filtering
in a nonlinear type-II BBO crystal \cite{yk2003}, or without any
post-selection or filtering just by controlling temperature in periodic
polarized nonlinear crystals (ppKTP, ppLN, etc.) \cite{rrf2009,zlz2024}. In
other word, if you want to manipulate the frequency degree of freedom, you
need to do post-selection using an optical element (e.g. single modal fiber)
or a filter with a certain wavelength width after SPDC in BBO, or control
temperature when using ppKTP. However, this kind of manipulation is
discontinuous. Even if it is temperature controls for ppKTP, large temperature
control can only get a small frequency manipulation range. In our method, the
key technology lies in the frequency modification by two rotational
meta-surfaces (q-plates) which are synchronously controlled by a computer.
This frequency modification is derived from RDE which has been widely
investigated in q-plate, rough surface and particles
\cite{sfm2018,yk2003,qcz2023,ljr2010,ss2013}. Especially for a rotating
q-plate, the incident circularly polarized light undergoes not only an energy
exchange through the rotational Doppler effect---causing the local electric
field vector of the circular polarization to rotate over time at the optical
frequency---but also a momentum exchange between SAM and OAM \cite{gsz2017}.
Thus, the entanglement is generated between the SAM, the OAM and frequency
degrees of freedom as
\begin{equation}
\psi_{h}=\frac{1}{\sqrt{2}}\left(  \left\vert \sigma^{-},+l,\omega
_{1}\right\rangle \left\vert \sigma^{+},-l,\omega_{2}\right\rangle +\left\vert
\sigma^{+},-l,\omega_{2}\right\rangle \left\vert \sigma^{-},+l,\omega
_{1}\right\rangle \right)  .
\end{equation}
The conversion processes are exhibited in Fig. 1. Here, $\omega_{1}%
=\omega+l\Omega$ and $\omega_{2}=\omega-l\Omega$ ($\Omega$ is the rotational
frequency of the media), corresponding to photons with OAM states $\left\vert
+l\right\rangle $ and $\left\vert -l\right\rangle $ respectively, are the
shifted frequencies after the rotational Doppler effect.

We can further convert the SAM state into a polarized state and remove the
effect of polarization with two polarizers and two QWPs, as seen in Fig. 1.
Then, a hybridized OAM-frequency entanglement state is formed%
\begin{equation}
\psi_{o}=\frac{1}{\sqrt{2}}\left(  \left\vert +l,\omega_{1}\right\rangle
\left\vert -l,\omega_{2}\right\rangle +\left\vert -l,\omega_{2}\right\rangle
\left\vert +l,\omega_{1}\right\rangle \right)  , \label{oss}%
\end{equation}
This state can be analyzed by two-photon interference at a fiber beam-splitter
(FBS) and by the coincidence counting.

\begin{figure}[ptb]
\centering
\includegraphics[width=0.9\columnwidth]{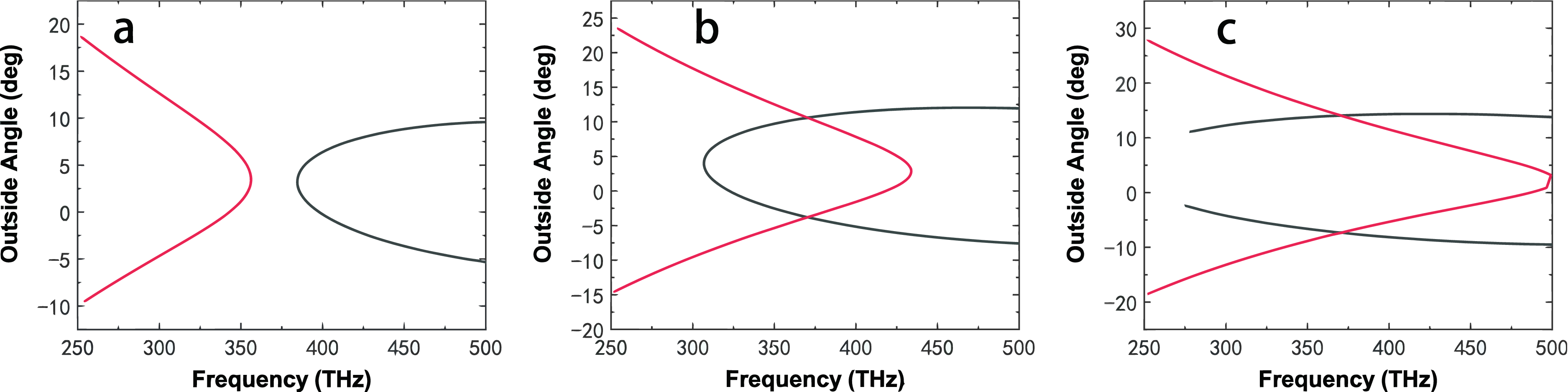} \caption{(Color online) From
left to right, we simulated the relation between the outside angle and the
light frequency in the spontaneous parametric down-conversion (SPDC) process
for different cut angles of 40${{}^{\circ}}$ (a), 45${{}^{\circ}}$ (b), and
50${{}^{\circ}}$ (c), respectively. In each diagram, the gray line represents
the ordinarily polarized light emitted from the BBO, while the red line
represents the extraordinarily polarized light. The intersection points of the
two curves indicate the frequency and emission angles of the entangled
photons.}%
\label{Fig2}%
\end{figure}

In our proposed experiment, the spectral width of the semiconductor
continuous-wave laser is typically around $1$ nm, which necessitates the use
of a high-speed rotating RqP to observe significant interference phenomena.
The absorption rates of the wave plates and lenses, as well as the coupling
efficiency of the coupler, can lead to a reduction in the overall detection
efficiency at the photon counters. To achieve efficient coupling, the position
of the signal photons can be pre-determined using a camera or calculated based
on phase-matching conditions. The phase-matching angle and the refractive
index of BBO can determine the outside angle, which is the angle between the
pump wave vector and the optical axis of the BBO crystal or the angle at which
entangled photons are emitted from the BBO crystal. The different outside
angles leads to the different emission directions for the signal photons. Fig.
2 present the relation between the outside angle and the light frequency for
different optic-axis cut angles of BBO crystals. It is seen that the ordinary
(o) and extraordinary (e) rays do not intersect for the cut angle of 40
degree. For the cut angles of $45$ and $50$ degrees, the intersection points
correspond to signal photon frequencies of $370.44$ THz and $370.32$ THz,
respectively. The signal photons generated by the BBO crystal initially
exhibit a frequency bandwidth of $\Delta f\approx0.08$ THz. For rotation
speeds on the order of THz, this results in an error of approximately $2\%$
when using a rotating q-plate with the topological charge $l=2$. However, when
the topological charge $l$ is increased to 10 or 100 or higher order, the
error will be reduced significantly.

\section{Results and Discussion}

In this section, we show the simulated joint spectral amplitude (JSA)
\cite{jw1997,dsw2013,dcb2016} of the hybrid OAM-frequency two-photon state of
Eq. (\ref{oss}) and the simulated HOM interference \cite{hom1987} in different
rotational conditions, to reveal and quantify the usage restrictions and
measurement conditions for such entangled states.

\begin{figure}[ptb]
\centering
\includegraphics[width=0.9\columnwidth]{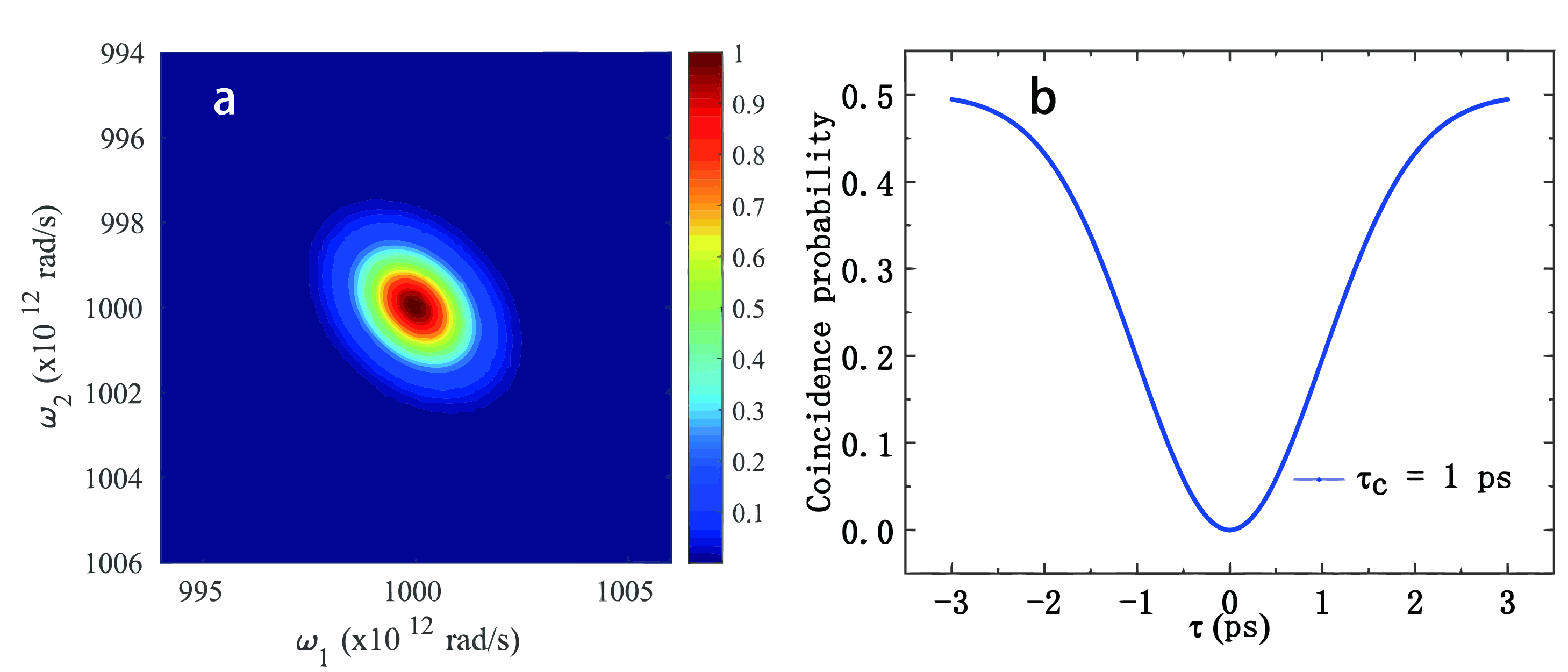} \caption{(Color online) (a)
Joint spectral amplitudes as a function of a signal frequency $\omega_{1}$ and
an idler frequency $\omega_{2}$, which is generated by laser beam going
through a BBO. (b) HOM coincidence pattern as a function of delay time for
polarization-entangled states. The parameters are taken as $\sigma=10^{12},
\gamma=0.1, A=-B=0.7/\sigma\sqrt{2\gamma}, \tau_{c}=1$ ps.}%
\label{Fig3}%
\end{figure}

In quantum optics, the JSA is a function that gives the associated amplitude
for the photon pair to have a signal frequency $\omega_{1}$ and an idler
frequency $\omega_{2}$. The JSA ($F\left(  \omega_{1},\omega_{2}\right)  $)
for these photons depend on a phase-matching function with a Gaussian profile
($\Phi\left(  \omega_{1},\omega_{2}\right)  $ that is only to simplify the
calculation without influencing our main analyses for entanglement) and a
Gaussian pump amplitude function ($\rho\left(  \omega_{1}+\omega_{2}\right)
$),%
\begin{equation}
F\left(  \omega_{1},\omega_{2}\right)  =\Phi\left(  \omega_{1},\omega
_{2}\right)  \rho\left(  \omega_{1}+\omega_{2}\right)  , \label{jsa}%
\end{equation}
where $\Phi\left(  \omega_{1},\omega_{2}\right)  $ is expressed as
$\Phi\left(  \omega_{1},\omega_{2}\right)  =\exp[-\gamma(A\omega_{1}%
+B\omega_{2})^{2}]$. The parameter $\gamma$ is the coefficient of the Gaussian
profile. $A$ and $B$ are the phase-matching parameters, depending on the
length ($L$) of the nonlinear crystal and the difference between the emitted
wave vectors ($k_{s}^{^{\prime}},k_{i}^{^{\prime}}$) of the signal and idler
photons and the pump wave vector $k_{p}^{^{\prime}}$, and they satisfy $A=-B$.
$\rho\left(  \omega_{1}+\omega_{2}\right)  =\exp[-\left(  \omega_{1}%
+\omega_{2}-2\overline{\omega}\right)  ^{2}/2\sigma^{2}]$ is a Gaussian
profile for the pump laser, $\overline{\omega}$ is the central frequency of
the pump, and $\sigma$ defines the spectral width. To produce polarization
entangled states with the same frequency under a certain spectral width in Eq.
(\ref{ps}), we consider two photons ($\omega_{1}=\overline{\omega}%
-\omega,\omega_{2}=\omega-\overline{\omega}$) at the intersection of the SPDC
rings. It gives rise to one separate JSA via frequency spectrum of signal and
idler photons, as shown in Fig. 3a. Gaussian phase-matching function makes it
expressed as an elliptical Gaussian intensity distribution, with the long axis
inclining along the antidiagonal direction.

If guiding these two indistinguishable polarization-entangled photons through
a beam splitter (BS) and let them coincidence count in a retardant time $\tau
$, they behave in a well-known HOM interference. It can be understood as the
coincidence probability for entangled photons after going though a BS as a
function of time delay $\tau$,%
\begin{equation}
P_{\tau}=\frac{1}{2}-\frac{1}{2}\int d\omega f^{\ast}\left(  -\omega\right)
f\left(  \omega\right)  e^{2i\omega\tau}, \label{pd}%
\end{equation}
where $f\left(  \omega\right)  =D\Phi\left(  \overline{\omega}-\omega
,\omega-\overline{\omega}\right)  $, and $D$ is the normalization constant
such that $\int d\omega\left\vert f\left(  \omega\right)  \right\vert ^{2}=1$.

\begin{figure}[ptb]
\centering
\includegraphics[width=0.9\columnwidth]{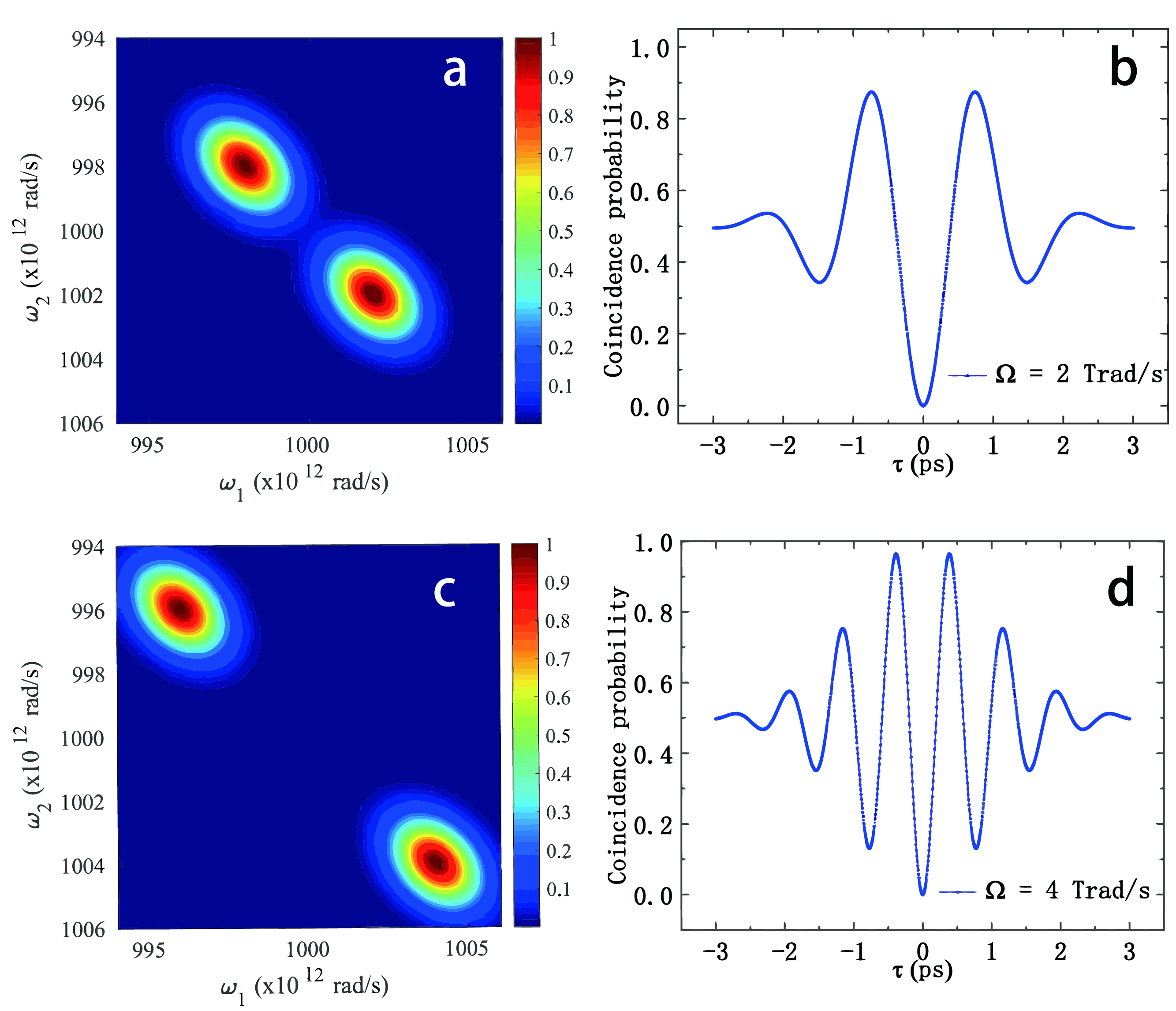} \caption{(Color online)
Joint spectral amplitudes (a,c) and HOM coincidence patterns (b,d) which are
similar as that in Fig.3, but the entangled state is replaced with
OAM-frequency-entangled states under the same topological charge $l=2$. Two
different rotational frequencies are taken as $2$ Trad/s ($1$ Trad/s =
$10^{12}$ rad/s) in (a,b) and $4$ Trad/s in (c,d). They have the same spectral
width given as $\tau_{c}=1$ ps.}%
\label{Fig4}%
\end{figure}

Under a pump spectral wide ($\Delta\omega=\sqrt{2}/\tau_{c}$) and a
phase-matching function with a Gaussian profile ($\Phi\left(  \overline
{\omega}-\omega,\omega-\overline{\omega}\right)  $), Eq. (\ref{pd}) becomes%
\begin{equation}
P_{\tau}=\frac{1}{2}-\frac{1}{2}e^{-\tau^{2}/\left(  2\tau_{c}^{2}\right)  },
\end{equation}
which exhibits a Gaussian dip in the interference plot as shown in Fig. 3b for
the state in Eq. (\ref{ps}).

After passing through a QWP and being modified by a rotational dielectric
metasurface (e. g. q-plate), the polarization-entangled photon pairs change
the entanglement states to Eq. (\ref{oss}). The phase match function
$\Phi\left(  \omega_{1},\omega_{2}\right)  $ changes to $\Phi\left(
\omega_{1}+l\Omega,\omega_{2}-l\Omega\right)  $, while the Gaussian profile
part $\rho\left(  \omega_{1}+\omega_{2}\right)  $ of JSA function maintains
the original form.

When the light goes through the rotating HWPs, the JSA shown in Fig. 3a
transfers to a new distribution in Fig. 4a for $l=2$ and $%
\Omega
=1$ Trad/s. We can clearly see that two separated peak patterns shift along
the antidiagonal direction. This offset will be more pronounced as the
rotational speed increase to $4$ Trad/s, as shown in Fig. 4c. This is due to
the frequency separation caused by the rotational Doppler effect. In our case,
the frequency $\omega_{1}+l\Omega$ is always associated to the $+l$ OAM state,
which is derived from the spin-flip $\left\vert \sigma^{+}\right\rangle
\rightarrow\left\vert \sigma^{-}\right\rangle $ and the angular momentum
conversation. As the SAM states $\left\vert \sigma^{+}\right\rangle $ arise
from $H$ polarization, the frequency $\omega+l%
\Omega
$ is always associated to the $H$ polarization, while the frequency $\omega-l%
\Omega
$ is always associated to the $V$ polarization. Because of tracing out the
polarization degree of freedom upon manipulation with simple optical elements,
there remains a composite OAM-frequency degree of freedom as the state in Eq.
(\ref{oss}). The OAM degree of freedom does not affect the frequency-frequency
entanglement and the corresponding JSA and HOM interferences.

When the signal and idler photons with frequency shift $\pm l\Omega$ are
delayed by a period of time $\tau$ and the output from a balanced BS with
modes $a$ and $b$, the entangle state transfers to%
\begin{equation}
\psi_{out}=\frac{1}{\sqrt{2}}\left(  e^{i\left(  \omega-l\Omega\right)  \tau
}\left\vert \omega-l\Omega\right\rangle _{a}\left\vert \omega+l\Omega
\right\rangle _{b}+e^{i\left(  \omega+l\Omega\right)  \tau}\left\vert
\omega+l\Omega\right\rangle _{a}\left\vert \omega-l\Omega\right\rangle
_{b}\right)  .
\end{equation}

And the coincidence probability between the beam-splitter outputs can be
calculated from the Gaussian phase-matching functions $\Phi_{+l}\left(
\omega+l\Omega\right)  $ and $\Phi_{-l}\left(  \omega-l\Omega\right)  $ of the
two photons,
\begin{equation}
P_{\tau}=\frac{1}{2}-\frac{1}{2}\cos\left(  2l\Omega\tau\right)  e^{-\tau
^{2}/\left(  2\tau_{c}^{2}\right)  },
\end{equation}
where the probability is obtained for $\left\vert \tau\right\vert <\frac
{\tau_{c}}{2}$, and the envelope width $\tau_{c}$ is related to the Gaussian
pump frequency band-width via $\tau_{c}=2\sqrt{2\ln2}/\Delta\omega
_{FWHM}\approx2.355/\Delta\omega_{FWHM}$. The degree of detuning frequency
$2l\Omega$ affects the frequency of oscillating signal in the HOM envelope
dig. If we want to see obvious intermediate jitters, the condition of
$2l\Omega>\Delta\omega_{FWHM}$ needs to be satisfied. When the envelope time
width $\tau_{c}=1$ ps, the corresponding $\Delta\omega_{FWHM}\approx1.4$
Trad/s, there will appear two cosine oscillations under the frequency shift of
$2l\Omega$ with $l=2$ and $\Omega=2$ Trad/s as shown in Fig. 4b. Obviously, if
increasing the rotational frequency $\Omega$ to $4$ Trad/s, this spatial
quantum beating (HOM interference) displays more oscillation as shown in Fig. 4d.

\begin{figure}[ptb]
\centering
\includegraphics[width=0.5\columnwidth]{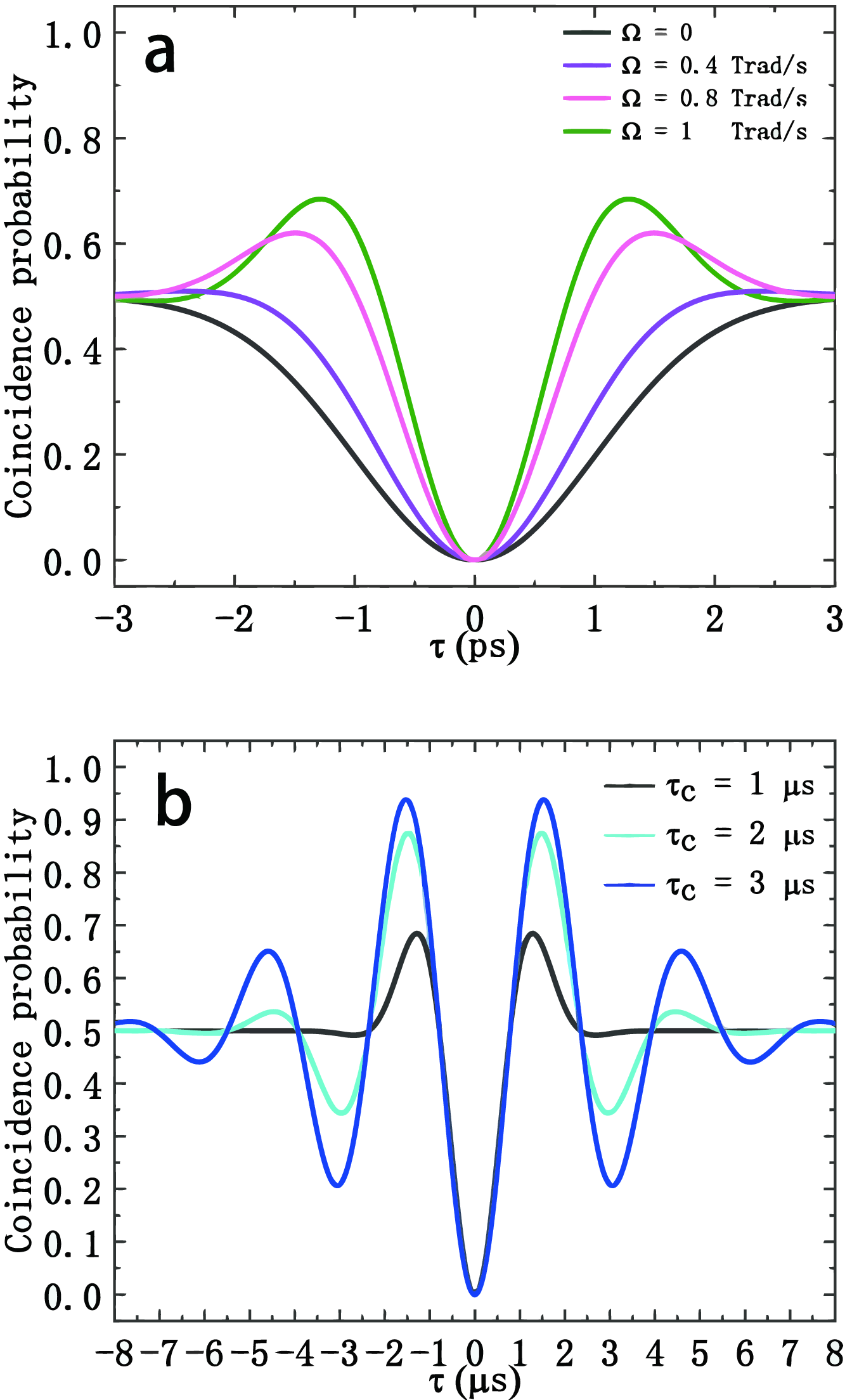} \caption{(Color online) HOM
coincidence patterns for frequency-entangled states under the same topological
charge $l=2$: (a) for different rotational frequencies 0, 0.4 Trad/s, 0.8
Trad/s, 1 Trad/s with the same spectral width $\tau_{c}=1$ ps; (b) for
different spectral width $1 \mu s, 2 \mu s, 3 \mu s$ but with the same
rotational frequency of $\Omega=1$ Mrad/s.}%
\label{Fig5}%
\end{figure}

Further reduction of rotational speed from $1$ Trad/s to $0.4$ Trad/s, the
intermediate jitters diminish rapidly to close to no detuning frequency state,
as shown in Fig. 5a. This result will lead to a challenge in our application
of quantum measurements for the two-photon frequency entanglement or the
rotational speed of the actual rotating body. For an mechanically-driven
rotating body, extreme speeds can be reached to GHz \cite{axb2018}, the
oscillations appear just when the envelope time width $\tau_{c}=1$ ns.
Correspondingly, the count time of the coincidence reaches to several
nanoseconds. In Fig. 5b, if we set the rotational frequency $%
\Omega
=1$ Mrad/s, the oscillations of the HOM interference obviously appear when
$\tau_{c}$ increase to $1$
$\mu$%
s and change more as the time width increases. Because $\tau_{c}$ is inversely
proportional to the width of the frequency spectrum of the Gaussian pump,
$\Delta\omega_{FWHM}=2\sqrt{2\ln2}/\tau_{c}\approx2.36$ Mrad/s, which leads to
a strict limitation of spectral width of pump light.

Finally, we would like to discuss, for lower rotation speed, what spectral
width can make HOM interference exhibit a significant and observable change
for a fixed delay time $\tau=1$ ps. For the case of no rotation, the
coincidence probability $P_{c}$ is approximately $0.2$ when the pump spectral
width $\tau_{c}=1$ ps. For the case of rotating the q-plate, the rotation
speed requires to reach $0.2$ Trad/s at least in order to observe a noticeable
change in HOM interference for the spectral bandwidth $\tau_{c}=1$ ps, taking
into the measurement noise. If we want to observe the HOM interference
obviously under the lower rotation speed, the spectral bandwidth will require
to become narrower (or $\tau_{c}$ will become larger). For example, if we take
$%
\Omega
=1$ Mrad/s, the obvious HOM interference pattern requires that $\tau_{c}$ is
approximately $1$ $\mu$s. When $\tau_{c}$ is less than $1$ $\mu$s (or the
frequency bandwidth is larger than $2.36$ Mrad/s), $P_{c}$ no longer exhibits
significant changes. Moreover, if the topological charge increases
significantly, an observable change in HOM interference can be achieved under
the current experimental conditions at a lower rotation speed or within a
lower bandwidth.

\section{Conclusion}

In this paper, we propose an experimental method for generating OAM-frequency
entangled two-photon states using the rotational Doppler effect. For these
states, we show that the joint spectral amplitude (JSA) and Hong-Ou-Mandel
(HOM) interference patterns differ significantly from those of
polarization-entangled states. In the case of HOM interference, we analyze the
appearance of oscillatory peaks in the interference dip, which are influenced
by the rotational frequency and the spectral bandwidth. When these two factors
are of comparable magnitude, distinct cosine-like oscillations emerge. Our
numerical simulations offer potential pathways for applications in quantum
information processing and quantum metrology.

\section*{Acknowledgments}

This work is supported by National Natural Science Foundation of China (NSFC)
with Grant No. 12375057, and the Fundamental Research Funds for the Central
Universities, China University of Geosciences (Wuhan).

\end{document}